\documentclass{PoS}

\title{Energy and Centrality Dependence of Chemical Freeze-out
 Parameters from Model Calculations}

\ShortTitle{Energy and centrality dependence of chemical freeze-out 
  parameters from models}

\author{\speaker{Lokesh Kumar}\\
        Kent State University, \\
        Kent, OH 44242\\
        E-mail: \email{lokesh@rcf.rhic.bnl.gov}}

\abstract{One of the main goals of heavy-ion collision experiments is
  to study the 
structure
of the QCD phase diagram. The QCD phase diagram is typically plotted as temperature
($T$) vs. baryon chemical potential ($\mu_{B}$). The statistical
thermal model THERMUS compared to experimental data provides
chemical freeze-out parameters 
such as temperature, baryon chemical potential and strangeness saturation factor
($\gamma_{s}$). However, the values of these parameters depend on models and their
underlying assumptions, such as the nature of the ensemble used, particle ratios
vs. particle yields, and the treatment of feed-down contributions to particle yields.
 In these proceedings, we report on a systematic study of chemical freeze-out
parameters using THERMUS, as a function of
collision centrality and collision energies ($\sqrt{s_{NN}} =7.7-200$ GeV). These
studies are performed with the string melting version of A Multi-Phase Transport
(AMPT) model. A comparison is presented of freeze-out parameters between
grand-canonical vs. strangeness canonical ensembles, particle yields vs. ratios,
with and without feed-down contributions to the particle yields. The main aim is to
evaluate the sensitivity of the thermal model fits to various model assumptions.
This is an important study for understanding corresponding experimental results from
the beam energy scan program at RHIC.}

\FullConference{8th International Workshop on Critical Point and Onset of Deconfinement,\\
		March 11 to 15, 2013\\
		Napa, California, USA}

\begin{document}

\section{Introduction}
One of the goals of heavy-ion collision experiments is to study the
QCD phase diagram. The QCD phase diagram is generally represented by
temperature ($T$) versus baryon chemical potential ($\mu_B$). Study of
hadron production plays an important role in understanding the
dynamics of relativistic collisions. The statistical thermal model has
been extremely successful in describing the hadron multiplicities
observed in these relativistic collisions. The statistical model or
THERMUS uses experimental yields (or ratios) as input and provides the
corresponding chemical freeze-out temperature and baryon chemical
potential~\cite{stm_therm}. 

Recently, the STAR experiment has presented the centrality dependence of the
freeze-out parameters using the Beam Energy Scan (BES) data. It is
observed that at lower energies ($\sqrt{s_{NN}}~\sim$7.7 and 11.5
GeV), the $T$ and $\mu_B$ vary with centrality~\cite{lokqm_sabqm}. Interestingly, for
Grand-Canonical Ensemble (GCE), the temperature seems to decrease
while going from central to peripheral collisions. On the other hand,
the temperature increases while going from central to peripheral
collisions when the Strangeness-Canonical Ensemble (SCE) is used. 
The baryon chemical potential decreases towards peripheral for both
GCE and SCE. 

In view of these observations, it is of interest to study the
centrality and energy dependence of freeze-out parameters using
transport models such as A Multi-Phase Transport
(AMPT) model~\cite{ampt}. For this study, the string-melting version of AMPT is
used. The particle yields ($dN/dy$) at mid-rapidity ($|y|<$0.5) are obtained for $\pi^{\pm}, K^{\pm},
p(\bar{p}), \Lambda(\bar{\Lambda}), K^0_S$, and $\Xi(\bar{\Xi})$ particles at
$\sqrt{s_{NN}}=$ 7.7, 11.5, 39, and 200 GeV. The errors on yields are
assumed to be of the order of 10\% for realistic comparison with experimental
results.
The centrality classes
used for this study are 0--5\%, 5--10\%, 10--20\%, 20--30\%, 30--40\%,
40--50\%, 50--60\%, 60--70\%, and 70--80\%. The THERMUS model is used
for extracting chemical freeze-out parameters using both
Grand-Canonical and Strangeness-Canonical ensembles. The freeze-out
parameters are extracted by using particle ratios as well as yields as
input to the THERMUS model.


\section{Results and Discussions}
\begin{figure}
\begin{center}
\includegraphics[scale=0.32]{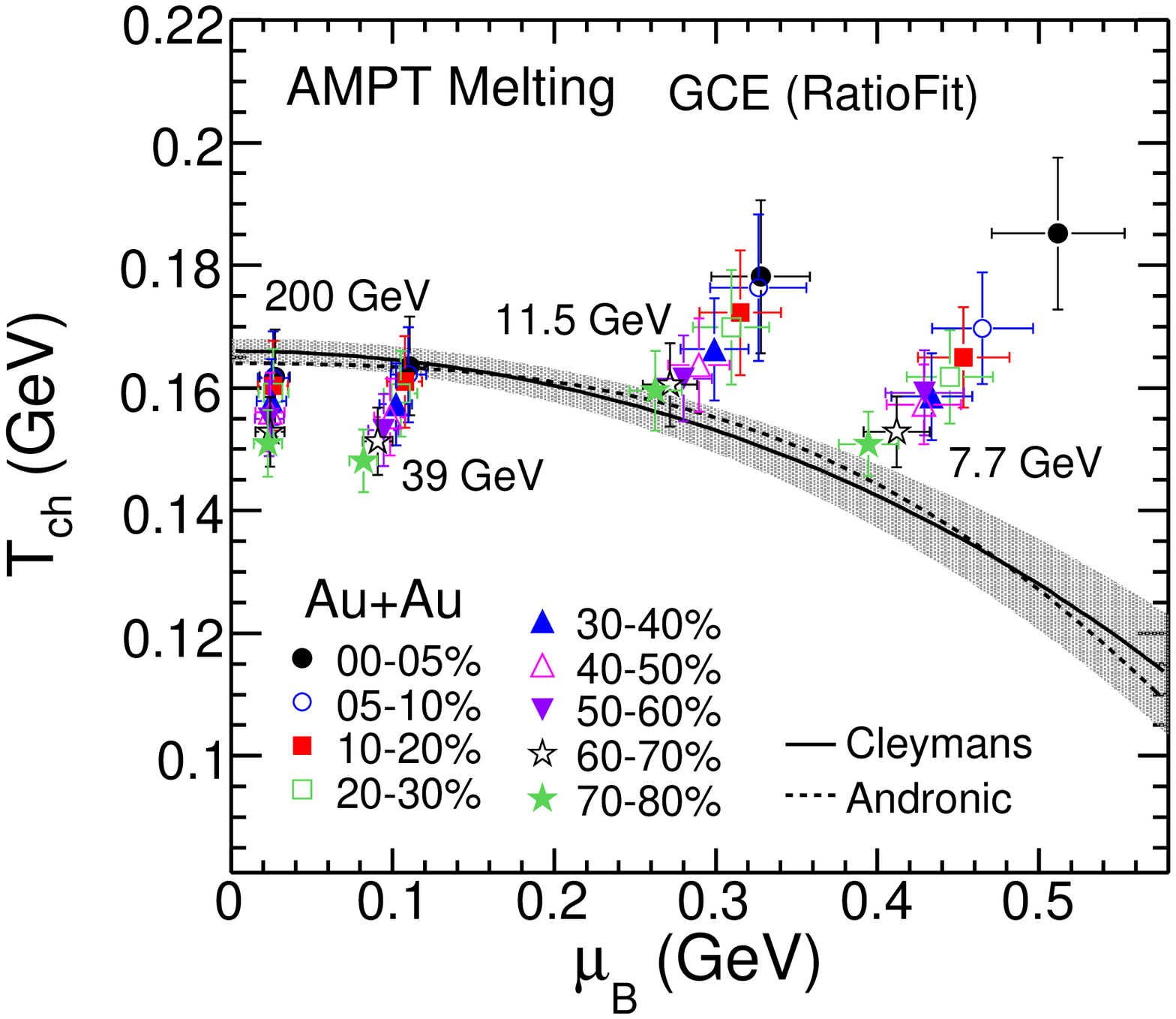}
\includegraphics[scale=0.32]{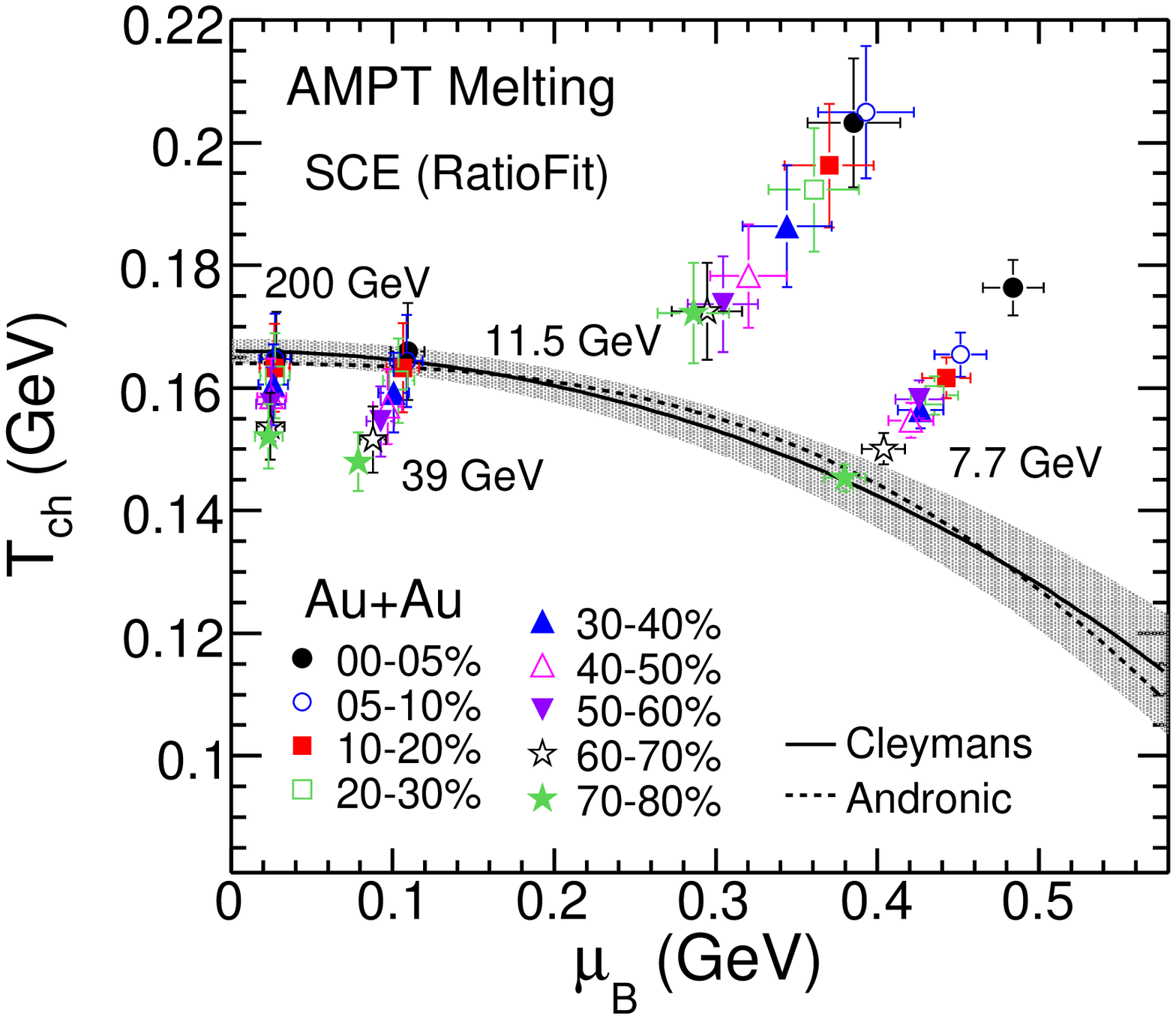}
\includegraphics[scale=0.32]{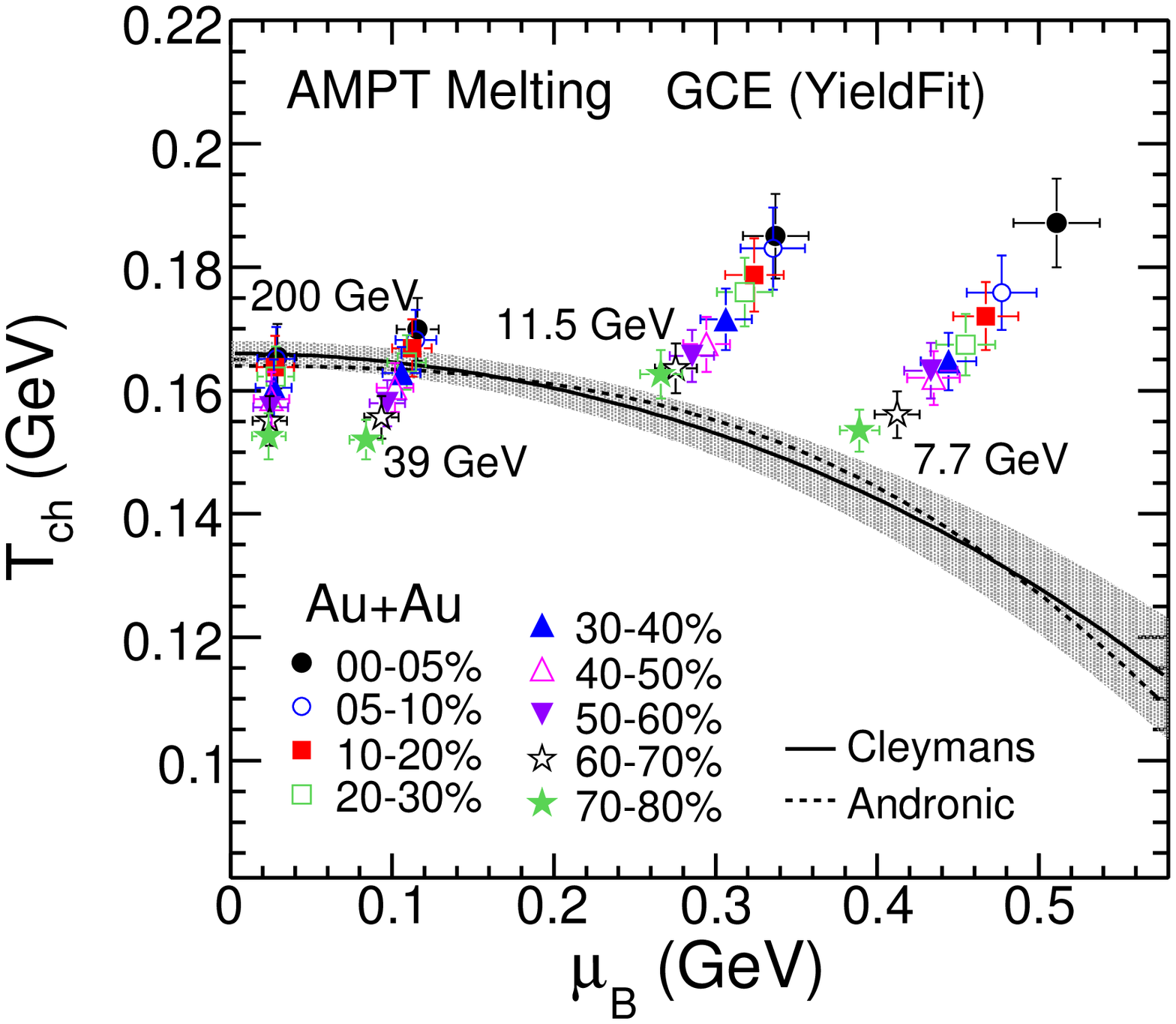}
\includegraphics[scale=0.32]{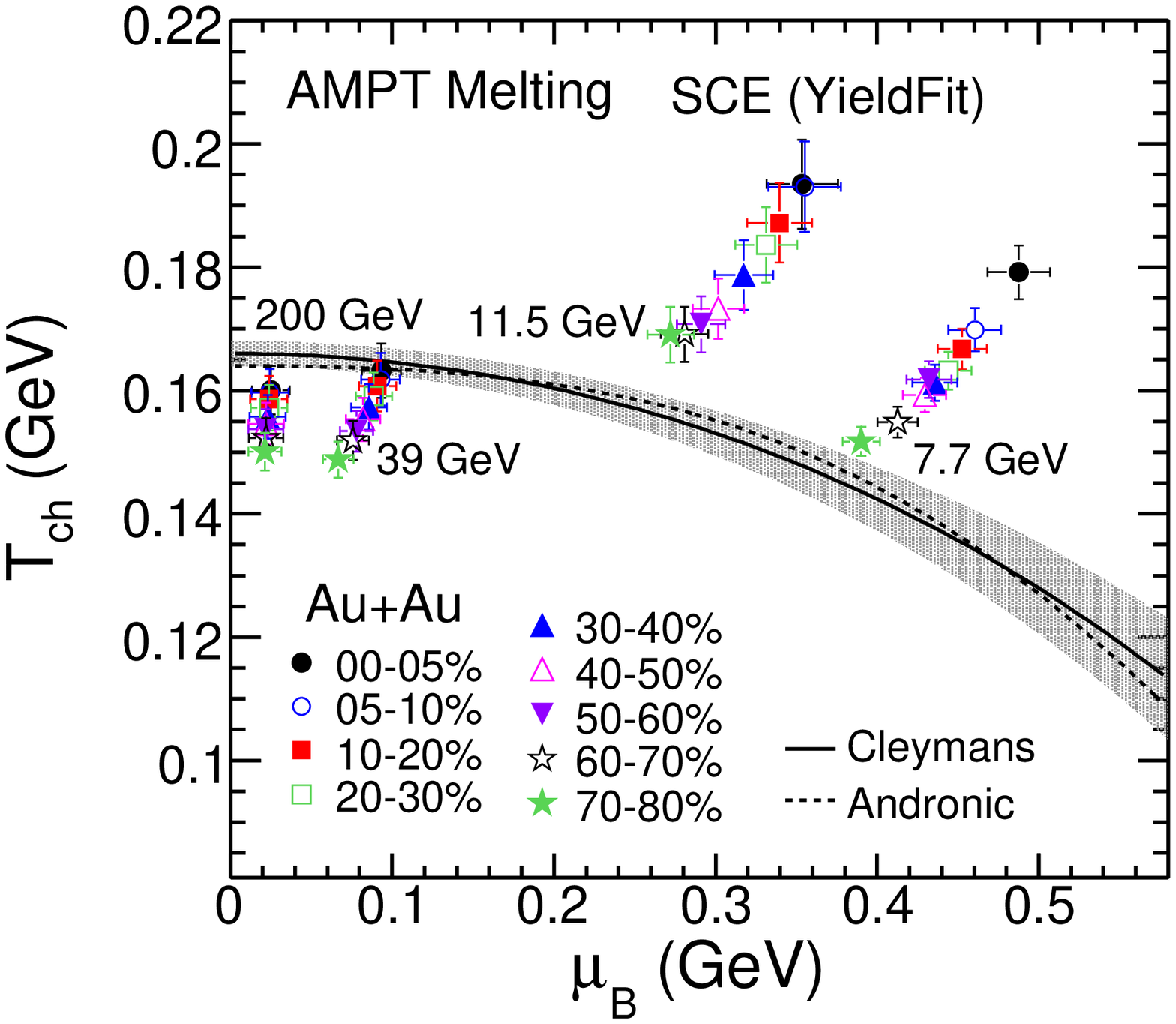}
\caption{ Top panels: Temperature vs. baryon chemical potential using
  particle ratios from AMPT as inputs to THERMUS model for GCE (left) and SCE
  (right). Bottom panels: Similar plots but obtained using particle
  yields from AMPT as inputs to THERMUS. Results are shown for
  beam energies 
from 7.7
  to 200 GeV and for different centralities.
}
\label{fig_tmub}
\end{center}
\end{figure}

Figure~\ref{fig_tmub} shows the variation of chemical freeze-out
temperature and baryon chemical potential for different energies from
7.7 to 200 GeV and different centralities. Top panels represent
results obtained by using particle ratios from AMPT as inputs to
the THERMUS model for GCE (left) and SCE (right) while bottom panels
represent results obtained by using particle  yields from AMPT as input to THERMUS.
Both $T$ and $\mu_B$ show similar behavior
as a function of centralities at all energies. Both decrease from
central to peripheral collisions for all the cases studied. This is
opposite to what was observed in the RHIC BES data, where for
Strangeness-Canonical Ensemble, the temperature increases from central
to peripheral collisions. 

The comparison of freeze-out parameters obtained by using different
inputs such as particle ratios and particle yield to THERMUS suggests 
that the temperature values are similar and lie within 5\% for these
cases. However, $\mu_B$
values could differ by a maximum of 20\% between the two cases for SCE. 
When different ensembles (GCE vs. SCE) are used, the temperature
values again lie within 5\% whereas $\mu_B$ differ by $\sim$20\%
between GCE and SCE if yields are used as input to THERMUS. 

\section{Feed-down Effect}
\begin{figure}
\begin{center}
\includegraphics[scale=0.36]{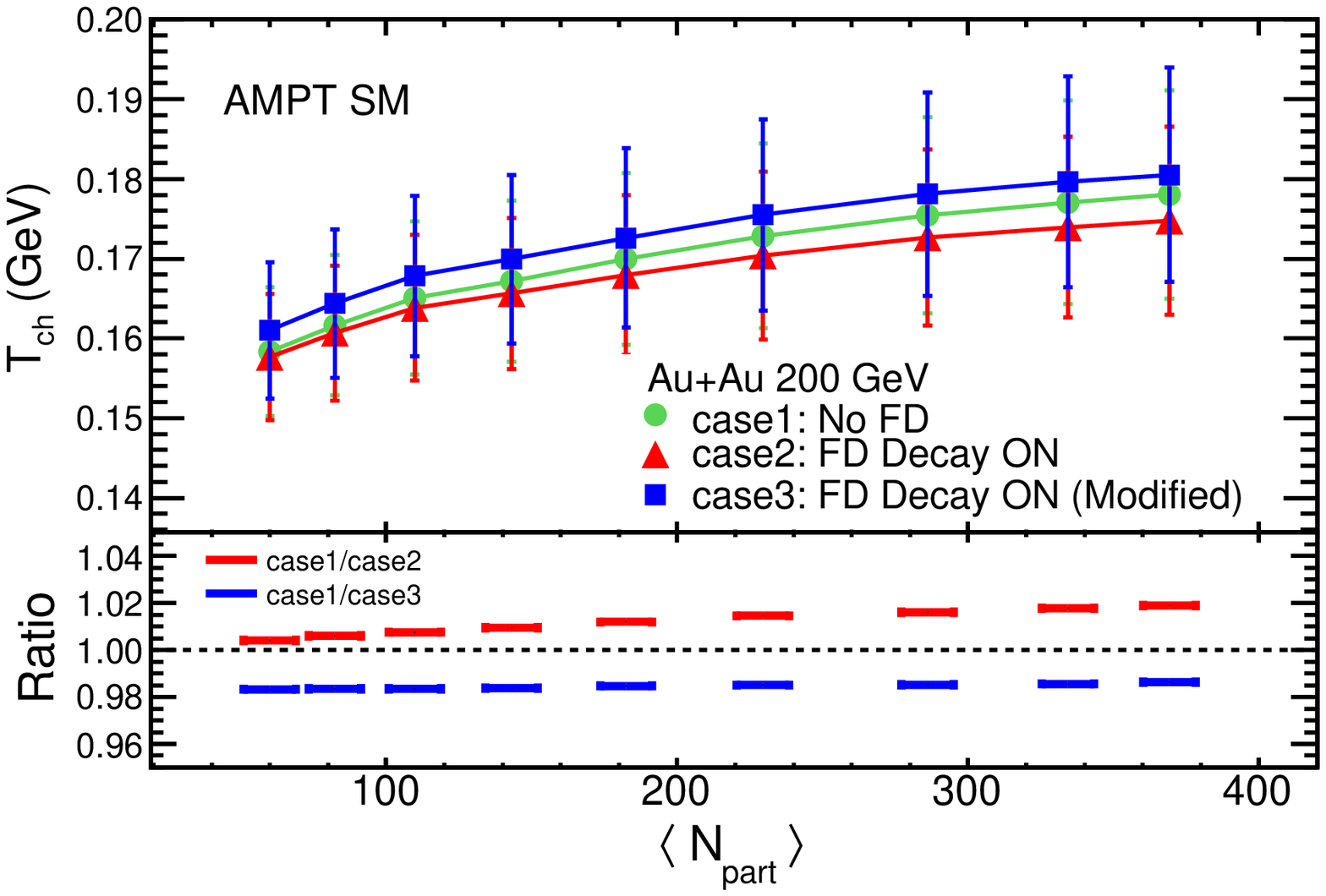}
\includegraphics[scale=0.36]{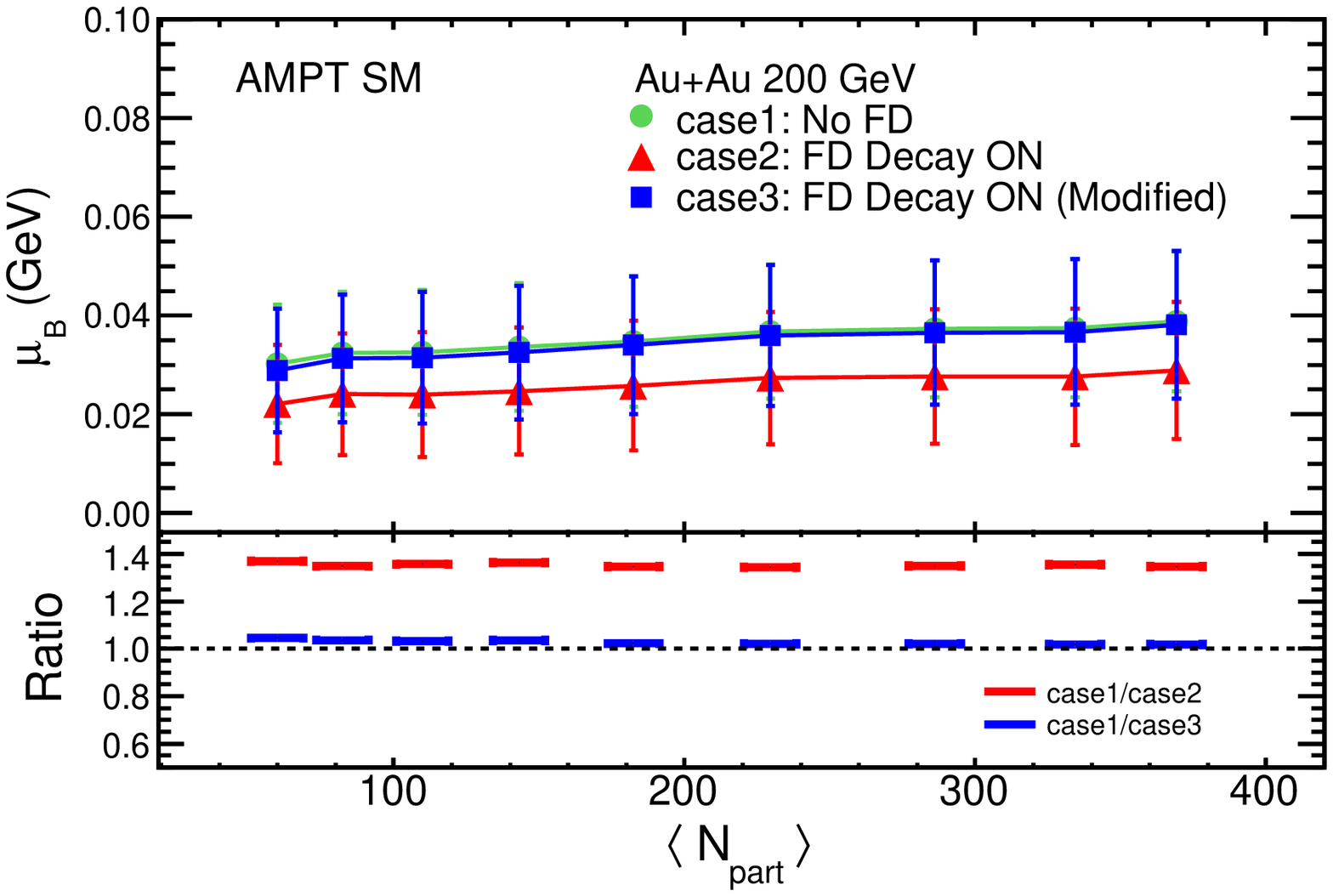}
\caption{ Top panels: Temperature (left) and baryon chemical
  potential (right) vs. $N_{\rm{part}}$  for three different cases in
  Au+Au 200 GeV using AMPT string melting. Bottom panels: Ratios of
  case1/case2 and case1/case3. See text for details.}
\label{fig_fd}
\end{center}
\end{figure}

In THERMUS, there is an option to switch OFF or ON the decay
channels. This option can be used to test the effect of feed-down 
contribution on the extracted freeze-out parameters. We consider
the weak-decay feed-down contribution from $\Lambda(\bar{\Lambda})$ to
$p(\bar{p})$. For this purpose, Au+Au 200 GeV from AMPT string melting is used. The $\Lambda$ and $\bar{\Lambda}$ are decayed
respectively to $p$ and $\bar{p}$. The Grand-Canonical approach is
used in THERMUS with input as the following particle ratios:
$\pi^-/\pi^+, K^-/K^+, \bar{p}/p$, and $\bar{p}/\pi^-$. The feed-down
contribution from $\Lambda$ to proton and $\bar{\Lambda}$ to $\bar{p}$
in THERMUS with default settings for Au+Au 200 GeV are 
of the order of 24\%. From AMPT, feed-down from $\Lambda$ to proton
is$~\sim$13\% while that from $\bar{\Lambda}$ to $\bar{p}$ is$~\sim$22\%. We study three cases: case 1 (No feed-down contribution
from $\Lambda(\bar{\Lambda})$ to $p(\bar{p})$ in both AMPT and
THERMUS), case 2 ( included feed-down contribution from
$\Lambda(\bar{\Lambda})$ to $p(\bar{p})$ both in AMPT and THERMUS),
and case 3 (feed-down contribution is included in both AMPT and
THERMUS but the contributions in THERMUS are modified so that both
AMPT and THERMUS have similar feed-down contributions from
$\Lambda(\bar{\Lambda})$ to $p(\bar{p})$). 
Figure~\ref{fig_fd} shows the results for the three cases for
temperature and baryon chemical potential vs. $N_{\rm{part}}$. Both
$T$ and $\mu_B$ show closer agreement for case 1 and case 3. The
$\mu_B$ values show large difference if the feed-down is not taken
care of properly in THERMUS (such as in case 2). 

\section{Summary}
We have presented the energy and centrality dependence of freeze-out
parameters from the AMPT model with a string melting
scenario. Freeze-out parameters are extracted using THERMUS and compared for
different ensembles (GCE vs. SCE) as well as for different inputs to 
THERMUS (particle yields vs. particle ratios). The $T$ and $\mu_B$
decrease from central to peripheral collisions for all
these cases in
contrast to what has been observed in STAR data, where $T$ increases
from central to peripheral collisions in SCE. For both GCE vs. SCE and
particle yields vs. ratios inputs, $T$ values lie within 5\% while $\mu_B$
may differ by a maximum of 20\%. The effect of weak decay feed-down
(from $\Lambda(\bar{\Lambda})$ to $p(\bar{p})$) on the
extracted freeze-out parameters is also presented. THERMUS can be
tuned to adjust the feed-down contributions according to data (AMPT in
this case). The difference in feed-down contribution between data
(AMPT in this case) and THERMUS 
is reflected in the extracted $\mu_B$ values.\\ 
\\
{\bf Acknowledgement:}  We would like to thank Prof. D. Keane,
N. Ming, Prof. B. Mohanty, Prof. N. Xu, and STAR Collaboration for the
Physics discussions. This work is supported by DOE
grant DE-FG02-89ER4053.


\begin{thebibliography}{99}
\bibitem{stm_therm} 
A. Andronic {\it et al.},
Nucl. Phys. A {\bf 772}, 167 (2006);
J. Cleymans {\it et al.},
Phys. Rev. C {\bf 73}, 034905 (2006);
F. Becattini {\it et al.},
Phys. Rev. C {\bf 73}, 044905 (2006);
J. Cleymans {\it et al.}, Comp. Phys. Comm. {\bf 180}, 84 (2009).

\bibitem{lokqm_sabqm} L. Kumar (for the STAR Collaboration) Nucl. Phys. A
  {\bf 904}, 256c (2013); S. Das (for the STAR Collaboration)
  Nucl. Phys. A {\bf 904},
  891c (2013). 
\bibitem{ampt} Z. W. Lin {\it et al.}, Phys. Rev. C {\bf 72}, 064901
  (2005); Z. W. Lin {\it et al.}, Phys. Rev. C {\bf 64}, 011902
  (2001); B. Zhang {\it et al.}, Phys. Rev.C {\bf 61}, 067901 (2000).


\end{thebibliography}
\end{document}